\documentclass[aps,prl,amsmath,amssymb,reprint,superscriptaddress,nofootinbib,floatfix]{revtex4-2}

\usepackage{mathtools}
\usepackage{microtype}
\usepackage{tikz}
\usetikzlibrary{arrows.meta}

\usepackage{graphicx}
\usepackage{dcolumn}
\usepackage{bm}
\usepackage[T1]{fontenc}
\usepackage{mathrsfs}
\usepackage{cancel}
\usepackage[dvipsnames]{xcolor}
\newcommand{\authormain}{Farshid Jafarpour}
\newcommand{\titlemain}{Optimal concentration and gradient sensing by perfect monitors}
\usepackage[bookmarks={false}, pdfauthor={\authormain}, pdftitle={\titlemain}]{hyperref}
\hypersetup{colorlinks=true, linkcolor=BrickRed, citecolor=Violet, filecolor=OliveGreen, urlcolor=RoyalBlue, filebordercolor={.8 .8 1}, urlbordercolor={.8 .8 0}}

\usepackage{orcidlink}
\newcommand{\mean}[1]{\left\langle#1\right\rangle}

\newcommand{\Var}[1]{\operatorname{Var}\left({#1}\right)}
\newcommand{\Cov}[1]{\operatorname{Cov}\!\left({#1}\right)}

\newcommand{\trans}{^{\mathsf T}}
\newcommand{\tr}{\operatorname{tr}}

\newcommand{\eq}[1]{Eq.~\eqref{eq:#1}}
\newcommand{\Eq}[1]{Equation~\eqref{eq:#1}}
\newcommand{\fig}[1]{Fig.~\ref{fig:#1}}

\newcommand{\apx}[1]{Appendix~\ref{app:#1}}

\newcommand{\tightmathspacing}[1]{%
    \begingroup
    \thinmuskip=0mu
    \medmuskip=1mu plus 1mu minus 1mu
    \thickmuskip=2mu plus 1mu minus 1mu
    {#1}%
    \endgroup%
}

\newcommand{\dd}{\mathrm{d}}
\newcommand{\br}{\bm r}

\newcommand{\bu}{\bm u}
\newcommand{\bg}{\bm g}
\newcommand{\bs}{\bm s}
\newcommand{\be}{\bm e}
\newcommand{\bw}{\mathbf w}
\newcommand{\w}{\mathrm w}
\newcommand{\bE}{\mathbf E}
\newcommand{\bp}{\mathbf p}
\newcommand{\bn}{\bm n}
\newcommand{\bI}{\bm I}

\newcommand{\E}{\mathcal E}
\newcommand{\capa}{\mathcal C_\Omega}
\newcommand{\pol}{\bm\alpha_\Omega}

\begin{document}

\title{\titlemain}

\author{Farshid Jafarpour\,\orcidlink{0000-0002-2441-0296}}
\email{f.jafarpour@uu.nl}

\affiliation{Institute for Theoretical Physics, Department of Physics, Utrecht University, Utrecht, Netherlands}
\affiliation{Centre for Complex Systems Studies, Utrecht University, Utrecht, Netherlands}

\date{\today}

\begin{abstract}
Cells often estimate chemical concentrations from only a few diffusing molecules, whose stochastic motion limits sensing precision. We consider a perfect monitoring instrument that records the instantaneous molecular density throughout a finite region without perturbing the concentration field or distinguishing molecular identities. The standard Berg-Purcell estimator weights all positions uniformly. Here we derive the optimal spatial weighting within a large class of unbiased estimators for both concentration and gradient sensing. Variance minimization maps to an electrostatic problem. Surprisingly, although the instrument monitors the entire volume, the optimal estimator assigns all weight to its boundary, reducing the uncertainty in both concentration and gradient sensing. We extend the results to arbitrary geometries and spatial dimensions.
\end{abstract}

\maketitle

\begin{figure*}[t]
    \centering
    \includegraphics[width=0.67\linewidth]{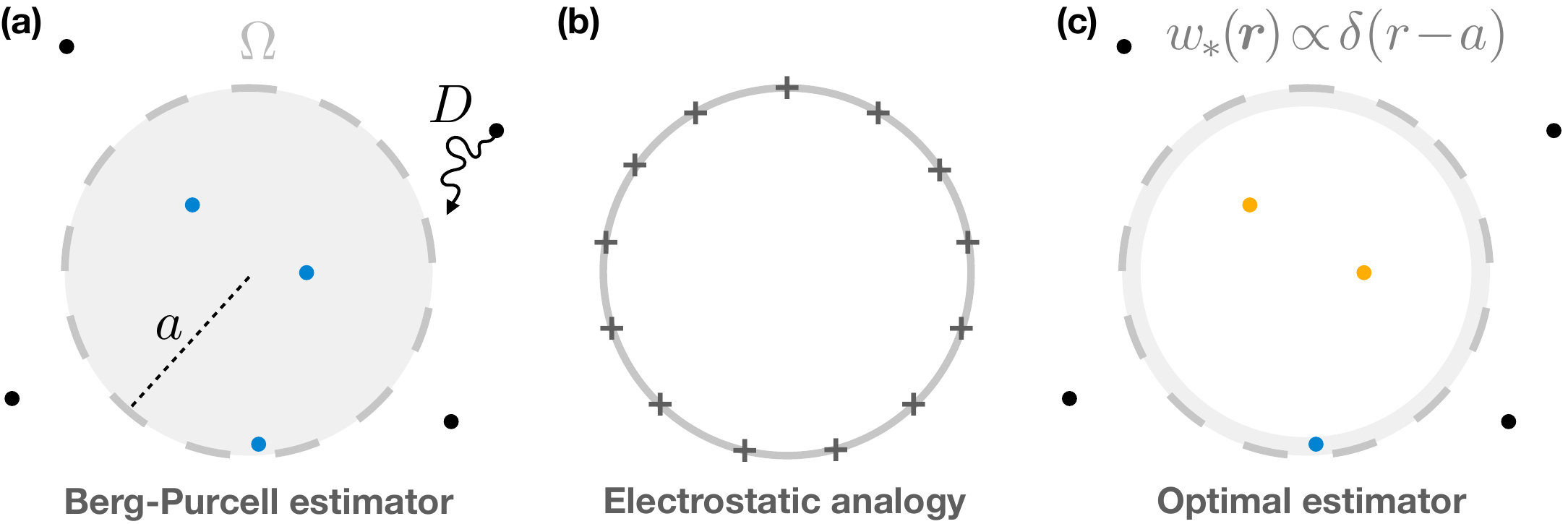}
    \caption{Schematic illustration of concentration sensing: (a) The classic Berg-Purcell instrument monitors a region $\Omega$ of space (in this case a sphere of radius $a$) and time averages the number of molecules within its volume (blue dots) as they diffuse in and out with a diffusion coefficient $D$. (b) Weight optimization maps to the electrostatic problem of charges on a conducting sphere, whose solution is a uniform charge distribution on the surface. (c) The optimal instrument time-averages the weighted count of molecules with weights given by the uniform surface charge density, thereby only counting the molecules within an infinitesimal shell near the boundary (blue dot) and ignoring molecules within its volume (orange dots).}
    \label{fig:concentration}
\end{figure*}

Bacterial cells can climb concentration gradients in environments so dilute that they encounter only a few molecules at a time. Motivated by this observation, Berg and Purcell showed in 1977 that the precision of cellular concentration sensing is fundamentally limited by the diffusive arrival and stochastic counting of molecules~\cite{berg1977physics}. Their work has motivated decades of research on the physical limits of cellular sensing~\cite{bialek2005physical,gregor2007probing,endres2008accuracy,bialek2008cooperativity,endres2009maximum,hu2010physical,mora2010limits,skoge2011dynamics,govern2012fundamental,mehta2012energetic,skoge2013chemical,kaizu2014berg,mora2015physical,ellison2016cell,fancher2017fundamental,varennes2017emergent,mora2019physical,fancher2020precision,vennettilli2021precision,mattingly2021escherichia,lefebre2024role,wu2025optimal,alonso2025local,mattingly2026coli, mugler2026cell}.

Berg and Purcell considered molecules with mean concentration $c_0$, each diffusing with diffusion coefficient $D$. They defined a ``perfect monitoring instrument'' as a permeable sphere of radius $a$ that can observe the molecules within its volume without disturbing them. They asked for the minimum uncertainty with which such an instrument could estimate $c_0$ over a measurement time $T$. The standard estimator uniformly averages the instantaneous number of molecules within the volume. Concretely, let
\begin{equation}
    \rho(\br,t)=\sum_{i}\delta\left(\br-\br_i(t)\right)
\end{equation} 
denote the empirical density of the molecules, where the position $\br_i(t)$ of each molecule performs Brownian motion. In terms of $\rho(\br,t)$, the standard Berg-Purcell estimator is
\begin{equation}\label{eq:BP_est}
    \widehat c_{_{\rm BP}} \equiv \frac{1}{T}\int_0^T \frac{1}{\mathcal V}\int_\Omega \rho(\br,t) \, \dd^3r\,\dd t,
\end{equation}
where $\mathcal V$ is the volume of the observation region $\Omega$ (see \fig{concentration}(a)). For long measurement times, $T\gg a^2/D$, Berg and Purcell found
\begin{equation}\label{eq:BP_var}
    \Var{\widehat c_{_{\rm BP}}} = \frac{3\,c_0}{5\pi DaT}.
\end{equation}
This estimator has commonly been interpreted as giving the best precision attainable by any physical measurement instrument that monitors molecules within its volume for a fixed time without disturbing the concentration field or distinguishing individual molecules~\cite{berg1977physics,endres2008accuracy,endres2009maximum,mora2010limits,govern2012fundamental}. Given that this estimator discards the positional information of the molecules, it is natural to ask if an instrument subject to these conditions can do better.

Here we define a family of position-weighted estimators for the concentration
\begin{equation}\label{eq:weighted_c}
    \widehat c\,[w] \equiv \frac{1}{T}\int_0^T \int_\Omega w(\br)\rho(\br,t) \, \dd^3r\,\dd t,
\end{equation}
subject to the constraint
\begin{equation}\label{eq:c_constr}
    \int_\Omega w(\br)\,\dd^3r = 1,
\end{equation}
and minimize their variance with respect to the weight distribution. We will show that optimizing the weight distribution is mathematically equivalent to letting the charge distribution in a conducting sphere relax to its minimal potential energy state with all the weight concentrated on the surface. In other words, a perfect instrument should ignore the molecules within its volume and only count the ones infinitesimally close to the surface. 

We then extend these results to gradient sensing. Consider an instrument attempting to estimate the gradient vector $\bg$ in a concentration profile of the form
\begin{equation}\label{eq:linear_profile}
    c(\br)\equiv\mean{\rho(\br,t)} = c_0 + \bg\cdot\br, \quad\text{with}\quad a|\bg|\ll c_0.
\end{equation}
We define a family of estimators for $\bg$ as
\begin{equation}\label{eq:weighted_g}
    \widehat \bg\,[\bw]=\frac{1}{T}\int_0^T \int_\Omega \bw(\br)\rho(\br,t) \, \dd^3r\,\dd t,
\end{equation}
and show that, with suitable constraints on the vector-valued weight distribution $\bw(\br)$, this is an unbiased estimator of the gradient $\bg$. Optimizing the variance of each component of $\widehat\bg$ over $\bw(\br)$ maps to the electrostatic problem of finding the charge distribution in a net-neutral conducting sphere in an external electric field in that direction, constrained to a fixed dipole moment. Again, the optimal configuration ignores the molecules in the bulk and considers only those infinitesimally close to the surface, thereby improving the previously proposed limit on gradient sensing of a perfect monitor~\cite{endres2008accuracy}, which corresponds to the case of $\bw(\br)\propto\br$. 

We then generalize these results to $d$ dimensions and nonspherical instruments. We show that shape dependence enters the precision calculation only through the capacitance and the polarizability tensor in concentration and gradient sensing, respectively; that spheres are the worst instruments of a given volume for either concentration or gradient sensing; and that optimization improves sensing more in higher dimensions. In $d\le2$, recurrent diffusion (repeated return of particles) maps to a divergence of the corresponding electrostatic potential, increasing concentration-sensing noise, while this divergent contribution cancels for gradient sensing. Additionally, we show that the optimal monitor has exactly twice the variance of Berg and Purcell's perfect absorber, independently of the instrument's shape, and that the larger, shape-dependent gap traditionally quoted between the two instruments overestimates the cost of remeasuring returning particles. We now derive these results, starting with concentration sensing in a spherical instrument.

\textit{Concentration Sensing.---} 
To find the optimal concentration estimator, we first find the variance of \eq{weighted_c} and then solve the variational problem for constrained optimization. For the estimator $\widehat c\,[w]$ to be unbiased, its expected value should be equal to $c_0$,
\begin{equation}
\begin{split}
    \mean{\widehat c\,[w]}&=\frac{1}{T}\int_0^T \int_\Omega w(\br)\overbrace{\mean{\rho(\br,t)}}^{c_0} \, \dd^3r\,\dd t\\
        &= c_0\int_\Omega w(\br) \, \dd^3r = c_0
\end{split}
\end{equation}
resulting in the constraint given in \eq{c_constr}. To compute its variance, let us first define
\begin{equation}
    \overline\rho_T(\br)\equiv\frac{1}{T}\int_0^T\rho(\br,t)\,\dd t.
\end{equation}
The covariance of $\overline\rho_T(\br)$ is given by (see \apx{mean_density_cov})
\begin{align}
    \!\! C_T(\br,\br') &\equiv \mean{\delta\overline\rho_T(\br) \delta\overline\rho_T(\br')} \notag\\
        &=\frac{2c_0}{T}\int_0^T \!\left(1-\frac{\tau}{T}\right)\, G(\br-\br',\tau)\,\dd\tau \label{eq:finite_time}\\
        &\simeq \frac{2c_0}{T} \!\! \int_0^\infty \!\!\! G(\br \!-\! \br'\!,\tau)\,\dd\tau \!=\! \frac{c_0}{2\pi DT}\frac{1}{|\br-\br'|}, \label{eq:coulombcov}
\end{align}
where $G(\br,\tau)\equiv e^{-|\br|^2/4D\tau}/(4\pi D\tau)^{3/2}$ is the diffusion propagator. The approximation from \eq{finite_time} to \eq{coulombcov} assumes $T\gg a^2/D$. The central idea is that the correlation function of the time-averaged density fluctuations is proportional to the time integral of the diffusion propagator, which is precisely the Green’s function of the Laplacian. This is the origin of the electrostatic mapping, which will extend to other dimensions and other geometries. Taking the variance of \eq{weighted_c} and substituting the covariance from \eq{coulombcov}, we find
\begin{equation}
\begin{split}
    \Var{\widehat c\,[w]} &= \frac{c_0}{2\pi DT}\int_\Omega\!\dd^3r\int_\Omega\!\dd^3r' \frac{w(\br)w(\br')}{|\br-\br'|}.
\end{split}
\end{equation}
Therefore, the optimization of variance is identical to minimizing the Coulomb self-energy of the charge density $w(\br)$ under the constraint of charge conservation \eq{c_constr}. This energy is minimized by the equilibrium charge distribution on a conductor, which lies entirely on its surface. The optimal concentration estimator therefore places all sensing weight at the boundary (see \fig{concentration}).

For a sphere, the minimizing weight is the uniform surface charge density
\begin{equation}\label{eq:copt}
    w_* = \arg\min_w \int_\Omega\!\!\dd^3r \!\!\int_\Omega\!\!\dd^3r' \frac{w(\br)w(\br')}{|\br-\br'|} = \frac{\delta(r-a)}{4\pi a^2},
\end{equation}
with variance (see \apx{variation_general})
\begin{equation}\label{eq:optimal_c_var}
    \Var{\widehat c_*}=\frac{c_0}{2\pi DaT}.
\end{equation}
The ideal surface expression is the limit of counting molecules in a permeable shell of thickness $\epsilon$ and dividing by its volume. Its instantaneous shot noise diverges as $1/\epsilon$, but its integrated correlation time vanishes as $a\epsilon/D$, leaving the finite result in \eq{optimal_c_var}. Compared with Eq.~\eqref{eq:BP_var},
$\Var{\widehat c_*}/\Var{\widehat c_{_{\rm BP}}}=5/6$. This is in agreement with the recent permeable-shell calculation of Ref.~\cite{mccusker2025physical}.

\begin{figure*}
    \centering
    \includegraphics[width=0.98\linewidth]{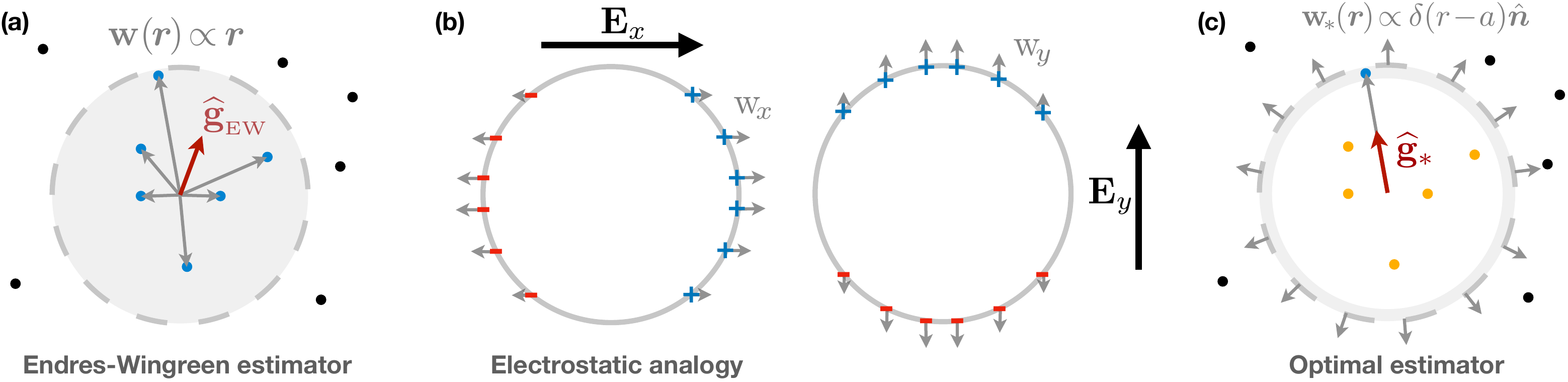}
    \caption{Schematic illustration of gradient sensing: (a) The Endres-Wingreen estimator $\widehat \bg_{_{\rm EW}}$ points toward the average position of molecules within its volume (blue dots). (b) The optimization problem maps each component of the estimator to the charge distribution of a net-neutral conducting sphere in an external electric field, with each component $\w_i$ of the weight vector field $\bw(\br)$ becoming the charge density in the corresponding electric field. (c) The optimal estimator points toward the average position of molecules within an infinitesimal shell near the boundary (blue dot) where the charge distributions are localized, ignoring the bulk molecules (orange dots).}
    \label{fig:gradient}
\end{figure*}

\emph{Gradient Sensing.---}
In 2008, Endres and Wingreen extended the Berg and Purcell results and defined a ``perfect monitoring instrument'' for estimating a weak concentration gradient $\bg$ of the linear concentration profile given in \eq{linear_profile}~\cite{endres2008accuracy}. In a given snapshot, the mean position of molecules in the instrument is expected to point in the direction of the gradient. Therefore, the Endres-Wingreen estimator $\widehat\bg_{_{\rm EW}}$ is defined to be proportional to the time average of the mean position of the molecules inside the region $\Omega$ (see \fig{gradient}(a))
\begin{equation}
    \widehat\bg_{_{\rm EW}} = \frac{15}{4\pi a^5}\frac1T\int_0^T\!\dd t\int_\Omega\!\dd^3r\, \br\rho(\br,t),
\end{equation}
where the prefactor ${15}/{4\pi a^5}$ is derived by setting $\mean{\widehat\bg_{_{\rm EW}}}=\bg$. The covariance of this estimator is given by
\begin{equation}\label{eq:EW_cov}
    \Cov{\widehat\bg_{_{\rm EW}}} = \frac{5\,c_0}{7\pi Da^3T}\bI,
\end{equation}
where $\bI$ is the identity matrix~\cite{endres2008accuracy}. The Endres-Wingreen estimator is a special case of the family of estimators in \eq{weighted_g} with the weight function proportional to the position vector $\bw(\br)\propto\br$. This estimator was originally derived by an ordinary least-squares fit of the time-averaged density to the linear profile. This procedure would minimize the variance if the density fluctuations were spatially uncorrelated; \eq{coulombcov} shows they are not.

Now let us consider the generalized family $\widehat \bg[\bw]$ defined in \eq{weighted_g}. For $\widehat \bg[\bw]$ to be unbiased we need 
\begin{equation}
\begin{split}
    \mean{\widehat \bg[\bw]} &= \frac{1}{T}\int_0^T \int_\Omega \bw(\br)\mean{\rho(\br,t)} \dd^3r\,\dd t \\
        &= \frac{1}{T}\int_0^T \int_\Omega \bw(\br)(c_0 +\bg\cdot\br)\, \dd^3r\,\dd t =\bg,
\end{split}
\end{equation}
which results in two constraints on $\bw(\br)$,
\begin{equation}\label{eq:constr_grad}
    \int_\Omega \bw(\br)\, \dd^3r = 0,\text{ and } \int_\Omega \w_i (\br)\,r_j \, \dd^3r = \delta_{ij}.
\end{equation}
Taking the covariance of \eq{weighted_g} using \eq{coulombcov}, we have
\begin{equation}\label{eq:cov_g_w}
    \Cov{\widehat \bg[\bw]} =  \frac{c_0}{2\pi DT}\int_\Omega\!\dd^3r\int_\Omega\!\dd^3r' \frac{\bw(\br)\bw\trans\!(\br')}{|\br-\br'|}.
\end{equation}
Minimizing the variance of each component $g_i$ of the gradient maps to the problem of minimizing electrostatic self-energy of the charge distribution $\w_i(\br)$, but now subject to the constraints in \eq{constr_grad}. The first constraint enforces that our conducting sphere is charge neutral, while the second fixes its dipole moment along the $i$ direction. Physically, this means that for each component $i$, the optimal $\w_i(\br)$ is exactly the induced charge distribution on a conducting sphere placed in a constant external electric field oriented along the $i$-axis. While each solution $\w_i(\br)$ is a dipolar charge distribution aligned with its respective axis, the full vector weight distribution $\bw(\br)$, constructed by combining all three orthogonal components, is a radially symmetric vector field 
\begin{equation}
    \bw_*(\br) = \frac{3}{4\pi a^3}\delta(r-a)\, \hat \br.
\end{equation}
This is an induced surface charge indicating that an optimal instrument only averages the positions of the molecules within an infinitesimal shell near the surface (see \fig{gradient}). The corresponding covariance of the optimal estimator is given by (see \apx{variation_general})
\begin{equation}\label{eq:optimal_cov}
    \Cov{\widehat{\bg}_*}=\frac{c_0}{2\pi Da^3T}\bI.
\end{equation}
Comparing \eq{EW_cov} and \eq{optimal_cov}, the optimal surface estimator of the gradient improves the covariance relative to the Endres-Wingreen estimator by a factor $7/10$.

\emph{Nonspherical Geometries.---}
The electrostatic analogy extends to measurement instruments of arbitrary shapes. Consider a solid conductor with a simply connected smooth shape $\Omega$. For the concentration sensing case, the optimal solution minimizing the variance of the estimator \eq{weighted_c} subject to charge conservation \eq{c_constr} gives a weight distribution $w_*(\br)$ equal to the equilibrium surface charge density of the conductor $\Omega$ with unit charge,
\begin{equation}
    w_*(\br) = \sigma(\br)\,\delta_{\partial\Omega}(\br),
\end{equation}
where $\delta_{\partial\Omega}(\br)$ is the surface $\delta$-function defined such that $\int_\Omega \!f(\br)\delta_{\partial\Omega}(\br)\dd^3r=\oint_{\partial\Omega}\!f(\bs)\dd S$. The same equilibrium charge distribution was recently shown to maximize the total uptake flux of absorbing receptors on an arbitrarily shaped cell surface~\cite{wu2025optimal}. Here this distribution arises from minimizing variance instead of maximizing flux and for a \emph{nonperturbing} estimator as opposed to an absorber.

To find the surface charge density $\sigma(\br)$, consider the solution to Laplace's equation
\begin{equation}
    \nabla^2u=0, \quad u=1 \text{ on }\partial\Omega,\quad\text{ and } \lim_{|\br|\to\infty} u(\br)=0,
\end{equation}
outside of $\Omega$. Then the surface charge density is given by
\begin{equation}
    \sigma(\br) = -\frac{1}{\capa}\frac{\partial u(\br)}{\partial n},
\end{equation}
where $\partial u(\br)/\partial n$ is the derivative of $u$ in the direction of the normal vector to the surface, and the capacitance $\capa$ is defined as
\begin{equation}
    \capa=-\oint_{\partial\Omega}\frac{\partial u}{\partial n}\,\dd S.
\end{equation}
For reference, the capacitance of a sphere of radius $a$ is equal to $\mathcal C_{\rm sphere}=4\pi a$. The variance of the optimal estimator is then given by (see \apx{variation_general})
\begin{equation}\label{eq:var_cap}
    \Var{\widehat c_*} = \frac{2c_0}{DT\capa}.
\end{equation}
Plugging $\capa=4\pi a$ for a sphere recovers \eq{optimal_c_var}. For comparison, a rotationally symmetric ellipsoid with long semi-axis $a$, two shorter semi-axes $b$, and eccentricity $e\equiv\sqrt{1-b^2/a^2}$ has the capacitance $\mathcal C_{\rm ellipsoid}={4\pi a\,e}/{\operatorname{arctanh} e}$. Among shapes of fixed volume, the sphere has the smallest capacitance~\cite{polya1951isoperimetric}, and therefore is the worst instrument for concentration sensing in the long-time limit. There is no best shape, since $\mathcal C_\Omega$ grows without bound as an ellipsoid of fixed volume is elongated.

\Eq{var_cap} also permits a direct comparison with the second idealized instrument considered by Berg and Purcell~\cite{berg1977physics} and Endres and Wingreen~\cite{endres2008accuracy}: a perfect absorber that removes each molecule upon first contact. A nonperturbing monitor may record the same molecule many times as it diffuses into and out of the observation region, since it cannot distinguish a returning molecule from a new one. By contrast, an absorber eliminates this additional source of uncertainty by removing the molecule after its first encounter. A perfectly absorbing body of shape $\Omega$ captures molecules at the mean rate $D\capa\, c_0$, giving $\Var{\widehat c_{\rm abs}} = {c_0}/{DT\capa}$. Comparing this expression with \eq{var_cap} gives
\begin{equation}
    \Var{\widehat c_*} = 2 \Var{\widehat c_{\rm abs}}.
\end{equation}
Therefore, the actual cost of not perturbing the concentration field is exactly a factor of two in variance, independent of the instrument's shape. The larger, shape-dependent cost obtained by comparing the absorber with the original Berg-Purcell estimator overestimates this effect.

For gradient sensing, placing a neutral conductor in an external electric field $\bE$ induces a dipole moment $\bp=\pol\bE$, where $\pol$ is the polarizability tensor. For reference, the polarizability of a sphere of radius $a$ is $\bm \alpha_{\rm sphere}=4\pi a^3 \bI$. Now we need to solve the electrostatic problem subject to the two constraints in \eq{constr_grad}. The first is charge neutrality, while the second, for each component $\w_i$, fixes the dipole moment along direction $i$. We first solve the electrostatic problem with a constant external electric field at infinity in each direction to find the polarizability tensor $\pol$, then we use $\pol$ to find the external electric field $\bE^{(i)} = \pol^{-1} \be_i$ that induces the dipole moment $\bp=\be_i$ in direction $i$, and solve the electrostatic problem again with the new external field to find the surface charge density and therefore the optimal $\w_i$. The covariance of the optimal estimator is given by (see \apx{variation_general})
\begin{equation}\label{eq:cov_pol}
    \Cov{\widehat{\bg}_*} = \frac{2c_0}{DT}\pol^{-1}.
\end{equation}
The trace of $\pol^{-1}$ is the same for all ellipsoids of the same volume; elongating a sphere increases the gradient sensing precision in that direction, but reduces it in other directions, keeping the total mean square error (MSE) constant. Nonellipsoidal shapes have smaller total MSE and outperform spheres and ellipsoids of the same volume, since ${\rm MSE}\propto{\rm tr}(\pol^{-1})\leq 1/|\Omega|$ (see \apx{polarizability}) with equality if and only if $\Omega$ is an ellipsoid~\cite{colbrook2026sharp}.

\emph{Generalization to $d$-Dimensions.---}
In $d=1, 2$ dimensions, a particle that leaves the instrument will return to the instrument over and over again with probability one, whereas in $d>2$ it has a nonzero probability of escaping forever. We will see that this distinction determines whether the long-time approximation we used to go from \eq{finite_time} to \eqref{eq:coulombcov} is valid. For $d>2$, \eq{coulombcov} becomes
\begin{equation}\label{eq:coulomb_d}
    C_T^{(d)}(R) \! \simeq \! \frac{2c_0}{T} \!\! \int_0^\infty \!\!\! \frac{e^{\frac{-R^2}{4D\tau}}}{(4\pi D\tau)^{d/2}}\,\dd\tau \!=\! \frac{c_0\,\Gamma\!\left(\frac d2 \!-\! 1\right)} {2\pi^{d/2}DTR^{d-2}},
\end{equation}
where $R\equiv|\br-\br'|$. The $1/R^{d-2}$ form of the correlations is also the Green's function of Poisson's equation in $d$ dimensions. The preceding electrostatic argument therefore carries over directly. For a $d$-dimensional ball of radius $a$,
\begin{equation}\label{eq:optimal_c_d}
    \Var{\widehat c_*} = \frac{\Gamma\!\left(\frac d2-1\right)}{2\pi^{d/2}} \frac{c_0}{DTa^{d-2}}, \qquad d>2.
\end{equation}
Comparing to the Berg-Purcell estimator, we have
\begin{equation}\label{eq:concentration_improvement_d}
    \frac{\Var{\widehat c_*}}{\Var{\widehat c_{_{\rm BP}}}} = \frac{d+2}{2d},
\end{equation}
which reduces to $5/6$ in three dimensions. The optimization process is more beneficial in higher dimensions.

In $d\leq 2$, \eq{finite_time} cannot be approximated by \eq{coulombcov}, since the integral in \eq{coulombcov} does not converge. While for $d>2$, the integral of the diffusion propagator $\int_0^T  G(R, \tau)\,\dd\tau$ converges to a constant as $T\to\infty$, it increases with $T$ as $\ln(T)$ in $d=2$ and as $\sqrt{T}$ in $d=1$ dimension. The divergence is the recurrence anticipated earlier: $\int_0^T G(R,\tau)\,\dd\tau$ is the mean time spent, up to $T$, at distance $R$ from the starting point, which grows without bound when the walk is recurrent and saturates when the particle can escape. This divergence maps to the divergence of the Coulomb potential at large distances in $d\leq2$ and can therefore be isolated by shifting the reference potential (see \apx{dimensions}), giving
\begin{equation}\label{eq:recurrent_concentration_scaling}
    \Var{\widehat c_*}\sim
        \begin{cases} 
            c_0/\sqrt{DT}, & d=1,\\[3pt] c_0\ln(DT/a^2)/(DT), & d=2.
        \end{cases}
\end{equation}
The recurrent contribution is independent of position and is therefore shared by all normalized concentration weights. Boundary localization remains optimal, but in $d\leq2$ it changes only the subleading part of the variance.

Gradient sensing does not inherit the low-dimensional problem, and the recurrent effects cancel out under the charge-neutrality condition $\int \w_i(\br)\,\dd^dr=0$.
The optimal dipole surface estimator retains $T^{-1}$ scaling in every dimension with covariance (see \apx{dimensions})
\begin{equation}\label{eq:optimal_gradient_d}
    \Cov{\widehat{\bg}_*}=\frac{\Gamma(d/2)}{\pi^{d/2}}\frac{c_0}{DTa^d}\bI_d,
\end{equation}
where $\bI_d$ is the $d\times d$ identity matrix. Compared to the linear estimator, we have
\begin{equation}\label{eq:gradient_improvement_d}
    \Cov{\widehat{\bg}_*}=\frac{d+4}{2(d+2)}\Cov{\widehat{\bg}_{_{\rm EW}}}.
\end{equation}
As with concentration sensing, the optimization has a larger effect in higher dimensions.

\emph{Conclusion.---}
We have shown that the standard Berg and Purcell estimator does not make optimal use of the spatial information available to a perfect monitoring instrument. Surprisingly, although the instrument observes molecules throughout its volume, the optimal estimator assigns all weight to its boundary, where molecules contribute less temporally redundant information than in the bulk. This result follows from an electrostatic mapping: concentration and gradient sensing correspond, respectively, to monopole and dipole variational problems, with the optimal precision for arbitrary geometries determined by capacitance and polarizability. The extension to different dimensions reveals that recurrence affects concentration sensing but cancels out in gradient sensing. Taken together, our results show that limits attributed to diffusive molecular noise can also depend on how an instrument combines the spatial information contained in its observations.

Although questions of concentration and gradient sensing were originally motivated by biology, the results presented here concern only the physical limits of measurement instruments under the stated constraints, and not how sensing is implemented in any particular biological or engineered system.

\begin{acknowledgments}
    \textit{Acknowledgments.---} I thank Andrew Mugler for helpful discussions and thoughtful feedback on the manuscript. 
\end{acknowledgments}

\bibliographystyle{apsrev4-2}
\bibliography{ref}
\newpage~\newpage

\appendix

\renewcommand{\section}[1]{%
  \par\addvspace{0.5\baselineskip}%
  \refstepcounter{section}%
  \noindent\textbf{Appendix~\thesection: #1.\textemdash}%
  \nobreak\hspace{0.5em}\ignorespaces
}

\setcounter{secnumdepth}{1}

\section{Time-averaged density covariance}\label{app:mean_density_cov}
Consider a sufficiently large volume of space in which boundary effects can be neglected, and the concentration profile is maintained ($c(\br)=c_0$ for concentration sensing and $c(\br) = c_0+\bg\cdot\br$ for gradient sensing). For $t<t'$, the density correlator is
\begin{equation}\label{eq:A1}
\begin{split}
    \mean{\delta\rho(\br,t)\delta\rho(\br',t')} &\!=\! \mean{\rho(\br,t) \rho(\br',t')}-c(\br)c(\br')\\
        &\hspace{-1.9cm}\!=\! \sum_{i,j}\mean{\delta(\br\!-\!\br_i(t))\,\delta(\br'\!-\!\br_j(t'))} \!-\! c(\br)c(\br')\\[-1em]
        &\hspace{-1.9cm}\!=\! \overbrace{\sum_{i\neq j}\mean{\delta(\br\!-\!\br_i(t))}\mean{\delta(\br'\!-\!\br_j(t'))}}^{c(\br)c(\br')}\\
            &\hspace{-1.9cm}\quad+ \sum_{i}\mean{\delta(\br\!-\!\br_i(t))\,\delta(\br'\!-\!\br_i(t'))} \!-\! c(\br)c(\br')\\
        &\hspace{-1.9cm}\!=\! \sum_{i}\mean{\delta(\br\!-\!\br_i(t))\!\mean{\delta(\br'\!-\!\br_i(t'))|\br_i(t)}}\\
        &\hspace{-1.9cm}\!=\!c(\br)\,G(\br-\br',\,t'\!-\!t),
\end{split}
\end{equation}
where $G(\br,\tau)=\exp\left({{-|\br|^2}/{4D\tau}}\right)/{(4\pi D\tau)^{d/2}}$ is the diffusion propagator. 

The covariance of the time average is
\begin{align}\label{eq:A2}
\begin{split}
    C_T(\br,\br') &\equiv \mean{\delta\overline\rho_T(\br)\delta\overline\rho_T(\br')} \\
        &\hspace{-0.3em}= \frac{c(\br) \!+\! c(\br')}{2\,T^2} \!\! \int_0^T \!\!\! \int_0^T \!\! G(\br \!-\! \br',|t \!-\! t'|)\,\dd t\,\dd t'\\
        &\hspace{-0.3em}= \frac{c(\br) \!+\! c(\br')}{T^2} \!\!\int_0^T \!(T \!-\! \tau)\,G(\br \!-\! \br',\tau)\,\dd\tau.
\end{split}
\end{align}
For concentration sensing $c(\br)\equiv\mean{\rho(\br)}=c_0$ and \eq{A2} reduces to \eq{finite_time}. For gradient sensing $c(\br)\simeq c_0$ inside the detector due to the shallow gradient assumption $a|\bg|\ll c_0$, recovering the same equation.

\section{Variational solution and general geometry in $d>2$}\label{app:variation_general}
We would like to minimize the variance of the form 
\begin{equation}
\begin{split}\label{eq:var_general_app}
    \Var{\widehat c\,[w]} &= \int_\Omega\int_\Omega w(\br)w(\br')\, C_T(\br,\br')\, \dd^dr\, \dd^dr'.
\end{split}
\end{equation}
In dimensions $d>2$,
\begin{equation}
    C_T(\br,\br')\simeq \frac{2c_0}{T}\int_0^\infty\!\! G(\br-\br',\tau)\dd\tau=\frac{2c_0}{DT}G(\br-\br'),
\end{equation}
where $G(\br)\equiv D\int_0^\infty G(\br,\tau)\dd\tau$ is the $d$-dimensional Coulomb potential satisfying $-\nabla^2G=\delta(\br)$.
Let us define the energy functional 
\begin{equation}
    \E[w] \equiv \frac12\int_\Omega\!\dd^d r\int_\Omega\!\dd^d r'\, w(\br)w(\br')\,G(\br-\br').
\end{equation}
For concentration sensing, we would like to minimize the energy (the variance $\Var{\widehat c[w]}=4c_0\E[w]/DT$) subject to charge conservation. We introduce a Lagrange multiplier $\lambda$ and the Lagrangian
\begin{equation}
    \mathcal L_c[w] \equiv \E[w]-\lambda \left[ \int_\Omega w(\br)\,\dd^dr-1 \right].
\end{equation}
Let us define the electric potential
\begin{equation}
    \Phi_w(\br) = \int_\Omega w(\br')\,G(\br-\br')\,\dd^dr'.
\end{equation}
In terms of $\Phi_w$, the variation of the Lagrangian is
\begin{equation}
    \delta\mathcal L_c \!=\! \int_\Omega \!\! \delta w(\br) \!\left[\Phi_w(\br) \!-\! \lambda \right]\dd^dr \! =\!0 \implies \Phi_{w_*}\!(\br) \!=\! \lambda.
\end{equation}
The optimal charge distribution yields a constant potential inside $\Omega$. Taking a Laplacian, $\nabla^2\Phi_{w_*}(\br)=-w_*(\br)$, shows zero charge density in the interior of $\Omega$; it is the equilibrium charge distribution on $\partial\Omega$. 

Let $Q=\int_\Omega w(\br)\,\dd^dr$ be the total charge and $V=\lambda$ the constant potential of the equilibrium charge distribution. The capacitance of a conductor with shape $\Omega$ is defined as $\capa\equiv Q/V$. Because the potential is constant on the surface,
\begin{equation}
\begin{split}
    \E[w_*] &= \frac12\int_\Omega w_*(\br)\underbrace{\Phi_{w_*}(\br)}_{V}\,\dd^dr= \frac12 VQ = \frac{Q^2}{2\,\capa}.
\end{split}
\end{equation}
For concentration sensing $Q=1$, and therefore
\begin{equation}\label{eq:B2}
    \E[w_*] = \frac{1}{2\,\capa} \implies \Var{\widehat c_*}=\frac{2c_0}{DT\capa}.
\end{equation}
This recovers \eq{var_cap} and its special case \eq{optimal_c_var} for the sphere in $d=3$ with $\capa=4\pi a$.

For gradient sensing, the claim is that there exists an admissible vector-valued weight $\bw_*(\br)$ such that it minimizes $\Var{\bu\cdot\widehat{\bg}[\bw]}$ for every unit vector $\bu$. Defining the scalar weight $\w_{\bu}(\br)\equiv\bu\cdot\bw(\br)$, we have
\begin{align}\label{eq:var_g_gen_app}
    \Var{\bu\cdot\widehat\bg[\bw]} &= \bu\trans\Cov{\widehat\bg[\bw]}\bu\\
        &\hspace{-5em}=\!\frac{2c_0}{DT}\!\int_\Omega\!\!\dd^dr \! \int_\Omega\!\!\dd^dr'\w_{\bu}(\br)\w_{\bu}(\br')G(\br-\br')\!=\!\frac{4c_0}{DT}\E[\w_{\bu}]\notag.
\end{align}
The unbiasedness constraints in \eq{constr_grad} imply
\begin{equation}\label{eq:B3}
    \int_\Omega\w_{\bu}(\br)\,\dd^dr=0, \qquad \bp\equiv\int_\Omega\br\,\w_{\bu}(\br)\,\dd^dr=\bu.
\end{equation}
We therefore minimize $\E[\w_{\bu}]$ subject to zero total charge and fixed dipole moment using the Lagrangian
\tightmathspacing{%
\begin{equation}
    \mathcal L_{\bu}[\w_{\bu}] \equiv \E[\w_{\bu}]-\lambda\int_\Omega\w_{\bu}(\br)\,\dd^dr-\bE\cdot\left[\int_\Omega\br\,\w_{\bu}(\br)\,\dd^dr-\bu\right]\notag.
\end{equation}}
Its variation gives
\begin{equation}
\begin{split}
    &\delta\mathcal L_{\bu}=\int_\Omega\delta\w_{\bu}(\br)\left[\Phi_{\w_{\bu}}(\br)-\lambda-\bE\cdot\br\right]\dd^dr=0 \\
    &\quad\implies\Phi_{\w_{\bu}^*}(\br)=\lambda+\bE\cdot\br.
\end{split}
\end{equation}
The total potential generated by $\w_{\bu}^*$ together with the external potential $-\bE\cdot\br$ is therefore constant inside $\Omega$. Thus, $\w_{\bu}^*$ is the charge induced on a neutral conductor in a uniform external electric field $\bE$, constrained to have dipole moment $\bp=\bu$. By definition of the polarizability tensor, $\bp=\pol\bE$, and therefore $\bE=\pol^{-1}\bu$. The minimum energy is
\begin{equation}\label{eq:B4}
\begin{split}
    \E[\w_{\bu}^*] &=\frac12\int_\Omega\w_{\bu}^*(\br)\Phi_{\w_{\bu}^*}(\br)\,\dd^dr\\
        &=\frac12\left[\lambda\int_\Omega\w_{\bu}^*(\br)\,\dd^dr+\bE\cdot\bp\right] \!=\! \frac12\bu\trans\pol^{-1}\bu.
\end{split}
\end{equation}
Consequently, every admissible $\bw$ satisfies
\begin{equation}\label{eq:gradient_lower_bound}
    \bu\trans\Cov{\widehat\bg[\bw]}\bu\geq\frac{2c_0}{DT}\bu\trans\pol^{-1}\bu.
\end{equation}

It remains to show that a single vector-valued weight attains this bound for every $\bu$. Let $\sigma_{\bE}(\br)$ denote the charge induced on a neutral conductor by an external field $\bE$, and define
\begin{equation}
    (\bw_*)_i(\br)\equiv\sigma_{\pol^{-1}\be_i}(\br)\, \delta_{\partial \Omega}(\br).
\end{equation}
We claim that this $\bw_*$ is the optimal vector-valued weight. The induced surface charge depends linearly on the external field. Therefore, for any $\bu=\sum_i u_i\be_i$,
\begin{equation}
    \bu\cdot\bw_*(\br)=\sum_i u_i\sigma_{\pol^{-1}\be_i}(\br)=\sigma_{\pol^{-1}\bu}(\br).
\end{equation}
The scalar weight $\bu\cdot\bw_*$ is the charge induced by the field $\pol^{-1}\bu$, which has dipole moment $\bp=\pol\pol^{-1}\bu=\bu$. It is therefore the minimizing weight for estimating $\bu\cdot\bg$, and equality holds in \eq{gradient_lower_bound} for every $\bu$:
\begin{equation}
    \bu\trans\Cov{\widehat\bg_*}\bu=\frac{2c_0}{DT}\bu\trans\pol^{-1}\bu.
\end{equation}
Equality of these quadratic forms for every $\bu$ gives
\begin{equation}
    \Cov{\widehat\bg_*}=\frac{2c_0}{DT}\pol^{-1},
\end{equation}
which recovers \eq{cov_pol} and its special case \eq{optimal_cov} for the sphere in $d=3$ with $\pol = 4\pi a^3\bI$.

\section{Bound on Trace of Inverse Polarizability}\label{app:polarizability}
Consider the weight function $\bw_n(\br) = \frac1{|\Omega|}\bn(\br) \delta_{\partial\Omega}(\br)$, where $|\Omega|$ is the volume of $\Omega$ and $\bn$ is the unit normal vector. It satisfies both conditions in \eq{constr_grad} and is therefore an admissible weight. The optimal weight $\bw_*(\br)$ would by definition outperform $\bw_n(\br)$ whose covariance is given by \eq{cov_g_w},
\begin{equation}
    \Cov{\widehat{\bg}_*} \!=\! \frac{2c_0}{DT}\pol^{-1} \preceq \Cov{\widehat{\bg}[\bw_n]} \!=\! \frac{2c_0}{DT}\mathbf B
\end{equation}
or $\pol^{-1} \preceq \mathbf B$, where $\mathbf B$ is defined as
\begin{equation}
    \mathbf B \equiv \frac{1}{|\Omega|^2}\int_{\partial\Omega}\!\!\!\dd S_{\br} \!\!\int_{\partial\Omega}\!\!\!\dd S_{\br'} \bn(\br)\bn\trans\!(\br')G(\br-\br'),
\end{equation}
with trace
\begin{equation}
\begin{split}
    \tr \mathbf B &\equiv \frac{1}{|\Omega|^2}\int_{\partial\Omega}\!\!\!\dd S_{\br} \!\!\int_{\partial\Omega}\!\!\!\dd S_{\br'} \bn(\br)\cdot\bn\!(\br')G(\br-\br')\\
        &=-\frac{1}{|\Omega|^2}\int_\Omega \dd^dr\int_\Omega \dd^dr' \nabla_{\br}^2G(\br-\br') = \frac{1}{|\Omega|}.
\end{split}
\end{equation}
Therefore, $\tr \pol^{-1}\leq 1/|\Omega|$. One can show by explicit calculations that for ellipsoids, $\bw_n$ is indeed the optimal weight, and therefore the bound is saturated for ellipsoids.

\section{Low Dimensions, $\bm{d\leq2}$}\label{app:dimensions}
For $d\leq2$, the approximation from \eq{finite_time} to \eq{coulombcov} does not hold, since the time integral of the diffusion propagator $\int_0^TG(R,\tau)\,\dd\tau$ does not converge as $T\to\infty$. We can still obtain a long-time limit that preserves the electrostatic analogy by using the fact that the integral $\int_0^T [G(R,\tau)-G(R_0,\tau)] \dd\tau$ converges as $T\to\infty$ for some reference distance $R_0$. This is done by adding and subtracting the term $G(R_0,\tau)$ under the integral in \eq{finite_time} and then taking the long-time limit, obtaining 
\begin{equation}
\begin{split}
    C_T(R) &\simeq C_T(R_0) \!+\!\! \frac{2c_0}{T}\!\!\! \int_0^\infty\!\!\!\! \left[G(R,\tau) \!-\! G(R_0,\tau)\right]\!\dd\tau\\
        &= C_T(R_0) + \frac{2c_0}{DT}G(R),
\end{split}
\end{equation}
where the new $G(R)\equiv D\int_0^\infty [G(R,\tau) \!-\! G(R_0,\tau)]\dd\tau$ still satisfies $-\nabla^2G=\delta(\br)$. This procedure is equivalent to shifting the zero-potential separation in the electrostatic analogy. For $d=1$, $G(R)$ is linear in $R$, and we can pick $R_0=0$, resulting in $G(R) = -R/2$. For $d=2$, $G(R)$ depends logarithmically, so $R_0$ has to be nonzero and finite (a natural choice for a circle of radius $a$ would be $R_0=a$, resulting in $G(R)=\ln(a/R)/2\pi$). For higher dimensions, we would pick $R_0=\infty$, which would remove the constant term. 

Using \eq{var_general_app}, we can write the variance as 
\begin{equation}\label{eq:var_c_app_shifted}
    \Var{\widehat c\,[w]} = C_T(R_0)Q^2 + \frac{4c_0}{DT}\E[w]+o(T^{-1})
\end{equation}
The optimization is not affected by the constant recurrence cost $C_T(R_0)$; the solution follows the same surface charge optimization and provides a constant potential $V$ on $\Omega$. The optimal energy is then $\E_*=VQ/2=V/2$, and
\begin{equation}
    \Var{\widehat c_*} = C_T(R_0) + \frac{2c_0V}{DT}+o(T^{-1}).
\end{equation}
This result holds in general dimension and for arbitrary shape $\Omega$, special cases of which are given in \eq{recurrent_concentration_scaling} for spheres in $d=1,2$.

The gradient sensing problem is simpler. As we saw in \eq{var_g_gen_app}, the variance of each component of $\widehat\bg$ has exactly the same form as the variance of $\widehat c$ (now in \eq{var_c_app_shifted}), but subject to a different set of constraints. Since the gradient case is charge neutral, the constant term in \eq{var_c_app_shifted} drops out. The optimal covariance is still given by $\Cov{\widehat \bg_*} = 2c_0/DT\, \pol^{-1}$ for any shape in any dimension. The special case of spheres in general dimension $d$ is given in \eq{optimal_gradient_d}.

\newpage

\end{document}